# Fiber-optic joint time and frequency transfer with the same wavelength

Jialiang Wang,[1,2] Chaolei Yue,[2,3] Yueli Xi,[3] Yanguang Sun,[3] Nan Cheng,[1] Fei Yang,[3*] Mingyu Jiang,[1,2] Jianfeng Sun,[2,3*] Youzhen Gui[1,2*], Haiwen Cai,[2,3]

[1] Key Laboratory for Quantum Optics, Shanghai Institute of Optics and Fine Mechanics, Chinese Academy of Sciences, Shanghai 201800, China
[2] Center of Materials Science and Optoelectronics Engineering, University of Chinese Academy of Sciences, Beijing 100049, China
[3] Key Laboratory for Space Laser Communication and Detection Technology, Shanghai Institute of Optics and Fine Mechanics, Chinese Academy of Sciences, Shanghai 201800, China
*Corresponding author: yzgui@siom.ac.cn；sunjianfengs@163.com; fyang@siom.ac.cn

**Optical fiber links have demonstrated their ability to transfer the ultra-stable clock signals. In this paper we propose and demonstrate a new scheme to transfer both time and radio frequency with the same wavelength based on coherent demodulation technique. Time signal is encoded as a binary phase-shift keying (BPSK) to the optical carrier using electro optic modulator (EOM) by phase modulation and makes sure the frequency signal free from interference with single pulse. The phase changes caused by the fluctuations of the transfer links are actively cancelled at local site by optical delay lines. Radio frequency with 1GHz and time signal with one pulse per second (1PPS) transmitted over a 110km fiber spools are obtained. The experimental results demonstrate that frequency instabilities of $1.7 \times 10^{-14}$ at 1s and $5.9 \times 10^{-17}$ at $10^4$s. Moreover, time interval transfer of 1PPS signal reaches sub-ps stability after 1000s. This scheme offers advantages with respect to reduce the channel in fiber network, and can keep time and frequency signal independent of each other.**

http://dx.doi.org/10.1364/OL.99.099999

The synchronization of time and frequency between remote locations is very important in many areas of human activity, involving Global Navigation Satellite System (GNSS) [1] as well as very long base line interferometry (VLBI) [2] and some commercial applications(e.g., metrology and timescale development). With the advances in modern atom clocks, such as cesium fountains, hydrogen masers and optical clocks, the frequency instability is better than 1E-13@1s for microwave signal [3, 4]. However the comparison of radio frequency signal using satellite-based techniques is limited at level around 1.0E-15 over one day. Optical fiber has provided a rapidly emerging alternative for long-distance time and frequency dissemination, offering much better accuracy.

Joint time and frequency transfer in optical fiber has been widely discussed and there are several kinds of schemes. One classical method for the purpose is transferring time and frequency signals in two adjacent wavelengths [5]. More specifically, time and frequency signals are modulated in light from two lasers which have difference wavelength about 0.8nm, two dense wavelength division multiplexers (DWDM) are used to combine and separate the modulated laser light at the local and remote sites. The dissemination stability is improved to 7.0E-15/1s and 4.5E-19/1day at the distance of 80km and time synchronization at the 50ps precision level was also demonstrated [6]. However, the communication channels in metropolitan network are scarce resources, this method may occupy an additional communicate channel. To get around this issue, scientists have proposed the other scheme, in which the sinusoidal frequency is converted into a square signal and the time signal (1PPS) is incorporated by triggering a violation made on the square wave by an embedder [7]. This scheme can solve the above problem, but the process of the sinusoidal frequency signal converting may introduce electronics noise from digital circuits. Those noises will cause a deterioration of frequency stability and make the short term stability worse than 2.0E-14/10s [8]. In the third scenario, mode-locked laser is employed to transfer time signal with intensity modulation, where time deviation 300fs@5s is obtained and optical and microwave frequency at a level better than 1.0E-17 are promising [9]. However, the technique is complicated because optical frequency combs are needed. In this letter, we propose a simple scheme to jointly transfer time and microwave frequency signal to the remote site with the same wavelength, whose major superiority is reducing the channel number in fiber network. At the local site, time signal is phase modulated and frequency signal is intensity modulated on the same carrier. At the remote site, the transferred optical carrier is split into two beams, time and frequency signals are recovered by coherent and incoherent demodulation, respectively. The noise caused by fiber link is compensated by the optical delay line. Time and frequency transfer over 110km fiber spools based on the proposed scheme is demonstrated. Relative stabilities of 1.7E-14/1s and 5.9E-17/10⁴s for frequency and time interval for 1PPS with sub-ps stability after 1000s are obtained. Furthermore, time signals in two locations can be synchronized with the method described in [7] and they can also be calibrated with Universal Time Coordinated (UTC).

Figure 1 illustrates the diagram of the proposed scheme. At the local site (LS), the reference time and frequency are modulated by the modulate system, which contains the laser source and two electro optical modulators (EOM). The modulation modes of two EOMs are different, one of which is based on intensity modulation (IM) and the other one is based on phase modulation (PM). The photoelectric field of optical carrier from laser source can be expressed as a simple E function

$$E_1 \propto \exp i(\omega_c t + \varphi_0), \quad (1)$$

where $\omega_c$ is the optical frequency of laser source and $\varphi_0$ is the initial phase.

The pulse amplitude is chosen in such a way that π-phase step is imprinted at the first EOM output. After that we obtain the photoelectric field

$$E_2 \propto \exp i[\omega_c t + \varphi_0 + \varphi(t)], \quad (2)$$

where $\varphi(t)$ is the encoded time signal which can be approximated as the rectangle function.

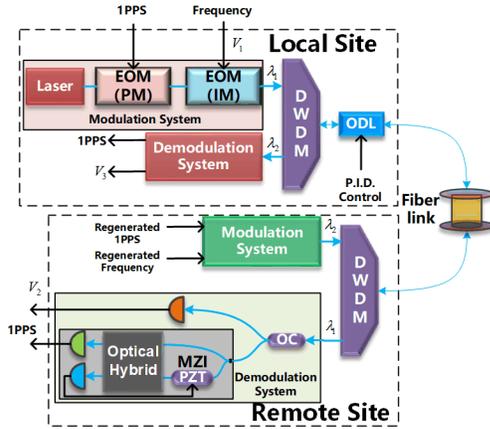

Figure 1. Schematic diagram of the proposed time and frequency transfer system. EOM, electro optical modulator; DWDM, dense wavelength division multiplexer; OC, optical coupler; ODL, optical delay line; PZT, piezoelectric transducer

The sine frequency signal can be expressed as

$$V_1 \propto \sin(\omega_0 t + \varphi'_0), \quad (3)$$

where $\omega_0$ is the frequency of sine signal and $\varphi'_0$ is the initial phase. The frequency signal is intensity modulated on the optical carrier by the second EOM. The light intensity after the 2nd EOM is proportional to

$$I_1 \propto [1 + \sin(\omega_0 t + \varphi'_0)]|\exp i[\omega_c t + \varphi_0 + \varphi(t)]|^2. \quad (4)$$

The optical carrier after the modulation system in the LS passes through the DWDM and injects into the optical delay line (ODL). The ODL contains a 10km temperature controlled fiber optic spool (bandwidth of several Hertz) and a short fiber wrapped around piezoelectrical transducer (PZT, bandwidth 2kHz) [10] in order to cancel out the slow and fast propagation delay caused by fiber link. After the ODL, forward signal passes the fiber link and arrives the remote site (RS) and its light intensity can be expressed as

$$I_2 \propto [1 + \sin(\omega_0 t + \varphi'_0 + \varphi'_{ODL} + \varphi'_1)]|\exp i[\omega_c t + \varphi_0 + \varphi_{ODL} + \varphi_1 + \varphi(t)]|^2, \quad (5)$$

where $\varphi'_1$ is the fiber link propagation delay from LS to RS for the RF signal and $\varphi'_{ODL}$ is the phase shift caused by ODL. $\varphi_1$ and $\varphi_{ODL}$ are the phase shift caused by fiber link and ODL for optical frequency, respectively.

The optical carrier arriving the RS can be recovered by the demodulation system which is shown in Figure 1 with light green block diagram. The light signal at the end of the fiber link is divided into two parts by a 50/50 optical coupler (OC). One part of which is inputted into a high-bandwidth photodetector (orange, Newport1611, bandwidth 30kHz-1GHz) and converted back to the electrical domain by direct detection (DD) of the intensity of received light. Ideally, pure phase modulation will not affect the intensity of the light signal [11]. After the detector we obtain frequency signal

$$V_2 \propto \sin(\omega_0 t + \varphi'_0 + \varphi'_{ODL} + \varphi'_1). \quad (6)$$

Then, the frequency signal is divided into two parts by a power splitter (shown in Figure 2). One is shown as regenerated frequency in the RS and remodulated back to the LS, the other is for the RS users.

The other part of the light signal from the OC is demodulated by a Mach-Zehnder interferometer (MZI), which is shown in gray block diagram. The MZI contains an optical hybrid, two photodetectors and one piezoelectric transducer (PZT, General Photonics, FPS-003-35-SS). This MZI has unbalanced paths and acts as delay line interferometer. After the optical hybrid, the phase difference between two unbalanced paths can be decoded by a low bandwidth photodetector (blue, Thorlabs, PDB140C). PZT is driven in such a way that path difference of two arms of the MZI equals as $\tau$ ($\tau$ is the width of the 1PPS signal, and also the fixed time delay produced by the unequal arm lengths of the MZI). A continuous wave will produce a destructive interference at the end of the MZI. The 1PPS pulse will produce by contrast two constructive interferences after recombination by the MZI [12-14]. The photoelectric fields of two beams inside the MZI can be expressed as

$$E_{MZI,1} \propto \sqrt{1 + \cos(\omega_0 t + \varepsilon)} \exp i[\omega_c t + \theta + \varphi(t)], \quad (7)$$

and

$$E_{MZI,2} \propto \sqrt{1 + \cos[\omega_0(t + \tau) + \varepsilon]} \exp i[\omega_c(t + \tau) + \theta + \varphi(t + \tau)], \quad (8)$$

where $\varepsilon$ and $\theta$ are the RF and optical phase before the MZI.

The light intensity after MZI can be described as

$$I_3 \propto 1 + \cos\frac{\omega_0 \tau}{2}\cos\frac{2\omega_0 \tau + 2\varepsilon + \omega_0 \tau}{2} + \left|\cos\frac{2\omega_0 \tau + 2\varepsilon + \omega_0 \tau}{2} + \cos\frac{\omega_0 \tau}{2}\right|\cos[\omega_c \tau + \varphi(t + \tau) - \varphi(t)], \quad (9)$$

It can be converted to electrical domain by DD, using a low bandwidth detector (green, Newport 1811, bandwidth DC-125MHz). This detector can act as a low-pass filter, it can filtrate the frequency signal. The time signal is recovered only after detector 1811 and input into a pulse distribution unit (PDU, Timetech No10188), whose outputs including the regenerated 1PPS and signal for users in the RS.

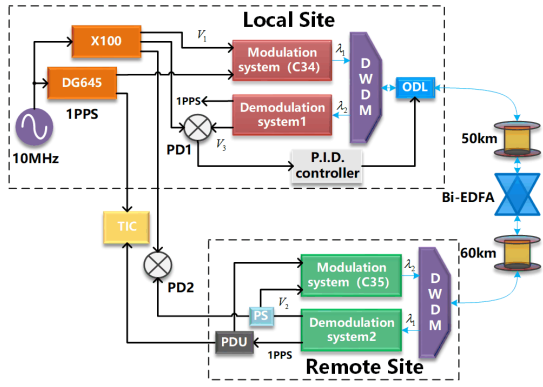

Figure 2. Experimental setup of the time and frequency transfer. PD, phase discriminator. PS, power splitter; PDU, pulse distribution unit.

The experimental setup is shown in Figure 2. A Rubidium atomic clock generates a 10MHz microwave signal and is split into two branches by a power splitter. One branch passes through a phase-lock frequency multiplier to obtain a 1GHz as the standard RF signal $V_1$. Meanwhile, the 1PPS time signal is produced by a time generator DG645, which is locked by the other frequency branch. Time and frequency signals are modulated on the laser in modulation system shown in figure 1 as light red block diagram. The optical carrier in LS has the wavelength $\lambda_1 = 1550.1nm$ (C34) and in RS has wavelength $\lambda_2 = 1549.3nm$ (C35). Different wavelengths are used to prevent Rayleigh scattering in the fiber spools and bi-directional erbium-doped fiber amplifier module (Bi-EDFA), the influence of bidirectional asymmetry can be neglected when the wavelength gap is small enough [15]. Two DWDMs are applied to separate the forward and backward optical carriers at the LS and the RS. The fiber link contains a 50km and a 60km fiber spool, between the two spools is a Bi-EDFA to amplify the amplitude of optical signals in both directions. At the RS, the signal carried on $\lambda_1$ is demodulated as shown in figure 1 and transmitted back along the same fiber link with the wavelength $\lambda_2$. The round-trip signal from the RS at the LS is demodulated by a same structure. The frequency signal in LS can be obtained as

$$V_3 \propto \sin(\omega_0 t + \varphi'_0 + 2\varphi'_{ODL} + 2\varphi'_1). \quad (10)$$

The signal $V_3$ and standard RF signal $V_1$ are input to a phase discriminator (PD1, HMC 439) to get the error signal [16]

$$\Delta\varphi_{err} = 2\varphi'_{ODL} + 2\varphi'_1. \quad (11)$$

After the process of proportional-integral-derivative (PID), we apply this error signal to drive the ODL, making it actuate a phase correction of $\varphi'_{ODL} = -\varphi'_1$. That means the whole fiber link is stabilized. Because the time and frequency carriers go through the same fiber path, the impact of vibration, temperature and humidity on the fiber are almost the same for the two signals. The time delay drift caused by fiber link is

$$t_l = \frac{\varphi'_1}{2\pi} \cdot T_{fre}, \quad (12)$$

where $T_{fre}$ is the period of frequency signal and the delay drift of ODL can be

$$t_{ODL} = \frac{\varphi'_{ODL}}{2\pi} \cdot T_{fre}. \quad (13)$$

When the whole fiber link is stabilized, $\varphi'_{ODL} = -\varphi'_1$ and $t_{ODL} = -t_l$. That is to say we can also obtain a stabilized time signal in the RS. The standard RF in the LS and the regenerated signal in the RS are input to the second phase discriminator (PD2) for the performance evaluation of frequency transfer. A time interval counter (TIC, GT668) is used to continuously measure the time delay between the regenerated 1PPS in the RS and that from the output of DG645.

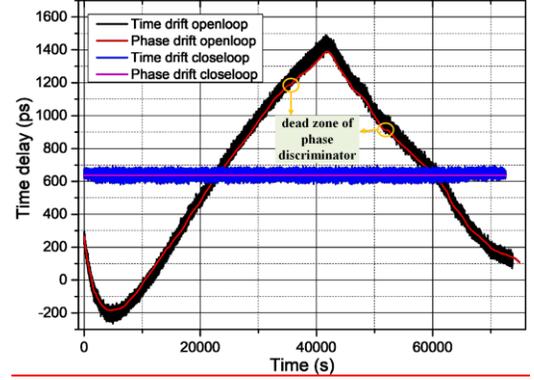

Figure 3. Time and phase drift between the RS and the LS.

Figure 3 shows the time delay drift and the frequency phase drift in the time-domain over 110km fiber link. The phase drift in open loop is mainly due to the temperature change, and the range of delay variation from fiber link is nearly 1800ps. It can be clearly observed that the time delay and frequency phase drift can fit well except that there is a tiny difference from 35000s to 52000s in open loop. We estimate this difference is due to the dead zone of phase discriminator because of drastic phase drift in open loop. For the case of closed loop, the fiber link is stabilized, and phase drift is limited in a small range, dead zone does not exist. The blue and magenta line show the time and phase drift for the case of close loop, the peak to peak value of the time delay drift is reduced to less than 120ps and the phase drift is reduced to less than 0.2ps.

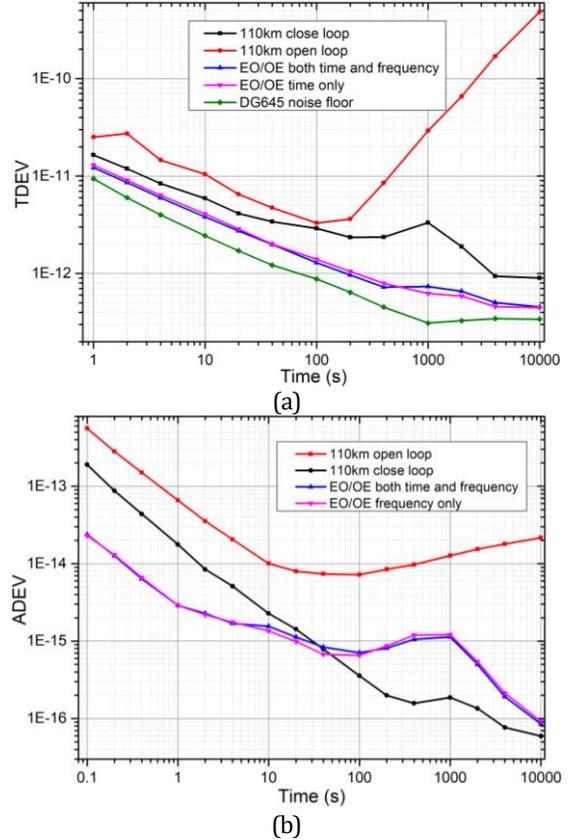

(a)

(b)

Figure 4. Measured time deviation (TDEV) (a) and Allan deviation (ADEV) (b) of the experiment scheme.

Figure 4(a) shows the measured time deviation (TDEV) of the proposed scheme over 110km fiber link. The red line is the result when the fiber link is running freely, it has short-term stability of 2.7E-11/1s and long-term stability worsens seriously because of temperature fluctuation. The dark line is the result when the phase noise is compensated, the users in RS can get a time signal with the relative stability of 1.6E-11/s and 9.1E-13/10$^4$s. There is a great improvement at averaging time of 100-10000s when comparing the performance of close loop with that of the open loop. Blue and magenta lines represent the TDEV noise from the electric-optic-electric (EO/OE) process when both time & frequency are modulated and only time is modulated, respectively. This EO/OE noise refers to free-running single trip and is obtained by connecting the LS and RS with a short fiber. The results from blue and magenta lines show that 1GHz frequency signal with intensity modulation has little impact on single pulse after the low-pass detector at the level of 10ps. At averaging time of 1000s, a periodic oscillation is obvious. The period of this oscillation coincides with that of our air conditioner, which can also be observed in frequency stability. The other limit of TDEV comes from the noise floor of DG645 (Green line, by measuring the time jitter between two outputs of DG645), which is about 9.4E-12/1s and 3.2E-13/10000s. The relative stability (Allan deviation, ADEV) of frequency transfer with a 5 Hz measurement bandwidth is shown in Figure 4(b). For the case of close loop over 110km fiber link, relative stability is 1.7E-14/1s and 5.9E-17/10$^4$s. Similarly, the EO/OE noise in two cases shown in Figure 4(b) indicate that the noise induced by residual amplitude modulation (RAM) [11] has a negligible impact on frequency stability at the level of fs.

In summary, we have proposed and demonstrated a fiber-based scheme that can transfer RF signal and single pulse with the same wavelength. Time and frequency signals are modulated in phase and intensity, respectively, so that they can be transferred with the same optical carrier. By this scheme, the channel resource can be saved in joint time and frequency transfer. A precise and continuous time and frequency transfer based on the proposed scheme is demonstrated over a 110km fiber link. Time and frequency signal are stabilized simultaneously owing to the ODL noise compensation. Results show the stability of frequency signal (ADEV) is 1.7E-14/1s and 5.9E-17/10$^4$s, and the TDEV of time signal is 1.6E-11/1s and 9.1E-13/10$^4$s. This scheme can be used for ultra-stable frequency dissemination and time synchronization system based on metropolitan network.

**Funding.** Strategic Priority Research Program of the Chinese Academy of Sciences under Grant XDB21030200 and in part by the National Natural Science Foundation of China under Grant 61535014

## References

1. Ashby, N. (2003). Relativity in the global positioning system. Living Reviews in relativity, 6(1), 1.
2. D. S. Robertson, Reviews of Modern Physics **63**, 899-918 (1991).
3. H. Marion, F. Pereira Dos Santos, D. Chambon, F. Narbonneau, M. Abgrall, I. Maksimovic, L. Cacciapuotti, C. Vian, J. Griinert, P. Rosenbusch, S. Bize, G. Santarelli, Ph. Laurent and A. Clairon," BNM-SYRTE fountains: Recent results", in 2004 Conference on Precision Electromagnetic Measurements. (IEEE,1976), pp. 495-495.
4. S. Weyers, V. Gerginov, M. Kazda, J. Rahm, B. Lipphardt, G. Dobrev, and K. Gibble, Metrologia **55**, 789-805 (2018).
5. W. Chen, Q. Liu, N. Cheng, D. Xu, F. Yang, Y. Gui, and H. Cai, IEEE Photonics Journal **7**, 1-9 (2015).
6. B. Wang, C. Gao, W. L. Chen, J. Miao, X. Zhu, Y. Bai, J. W. Zhang, Y. Y. Feng, T. C. Li, and L. J. Wang, Sci Rep **2**, 556 (2012).
7. P. Krehlik, Ł. Sliwczynski, Ł. Buczek, and M. Lipinski, IEEE Transactions on Instrumentation and Measurement **61**, 2844-2851 (2012).
8. Ł. Śliwczyński, P. Krehlik, A. Czubla, Ł. Buczek, and M. Lipiński, Metrologia **50**, 133-145 (2013).
9. M. Lessing, H. S. Margolis, C. T. A. Brown, and G. Marra, Applied Physics Letters **110** (2017).
10. O. Lopez, A. Amy-Klein, C. Daussy, C. Chardonnet, F. Narbonneau, M. Lours, and G. Santarelli, The European Physical Journal D 48, 35-41 (2008).
11. Whittaker, E. A., Gehrtz, M., & Bjorklund, G. C.,. JOSA B, **2**, 1320-1326(1985).
12. F. Frank, F. Stefani, P. Tuckey, and P. E. Pottie, IEEE Trans Ultrason Ferroelectr Freq Control **65**, 1001-1006 (2018).
13. Y. Zhi, J. Sun, Y. Zhou, W. Lu, E. Dai, W. Pan and L. Liu, Optical Engineering **58**, 016114 (2019).
14. A. M. J. van Eijk, Y. a. Zhi, J. Sun, E. Dai, Y. Zhou, L. Wang, W. Lu, P. Hou, L. Liu, C. C. Davis, S. M. Hammel, and A. K. Majumdar, "High-data rate differential phase shift keying receiver for satellite-to-ground optical communication link," presented at the Laser Communication and Propagation through the Atmosphere and Oceans2012.
15. L. Yu, R. Wang, L. Lu, Y. Zhu, J. Zheng, C. Wu, B. Zhang, and P. Wang, Opt Express **23**, 19783-19792 (2015).
16. J. Wang, W. Chen, Q. Liu, N. Cheng, Z. Feng, F. Yang, Y. Gui, and H. Cai, IEEE Photonics Journal **9**, 1-7 (2017).

1. Ashby, N. (2003). Relativity in the global positioning system. Living Reviews in relativity, 6(1), 1.
2. D. S. Robertson," Geophysical applications of very-long-baseline interferometry". Reviews of Modern Physics **63**, 899-918 (1991).
3. H. Marion, F. Pereira Dos Santos, D. Chambon, F. Narbonneau, M. Abgrall, I. Maksimovic, L. Cacciapuotti, C. Vian, J. Griinert, P. Rosenbusch, S. Bize, G. Santarelli, Ph. Laurent and A. Clairon," BNM-SYRTE fountains: Recent results", in 2004 Conference on Precision Electromagnetic Measurements. (IEEE,1976), pp. 495-495.
4. S. Weyers, V. Gerginov, M. Kazda, J. Rahm, B. Lipphardt, G. Dobrev, and K. Gibble," Advances in the accuracy, stability, and reliability of the PTB primaryfountain clocks". Metrologia **55**, 789-805 (2018).
5. W. Chen, Q. Liu, N. Cheng, D. Xu, F. Yang, Y. Gui, and H. Cai," Joint time and frequency dissemination network over delay-stabilized fiber optic links". IEEE Photonics Journal **7**, 1-9 (2015).
6. B. Wang, C. Gao, W. L. Chen, J. Miao, X. Zhu, Y. Bai, J. W. Zhang, Y. Y. Feng, T. C. Li, and L. J. Wang," Precise and continuous time and frequency synchronisation at the 5× 10-19 accuracy level". Sci Rep **2**, 556 (2012).
7. P. Krehlik, Ł. Sliwczynski, Ł. Buczek, and M. Lipinski," Fiber-optic joint time and frequency transfer with active stabilization of the propagation delay". IEEE Transactions on Instrumentation and Measurement **61**, 2844-2851 (2012).
8. Ł. Śliwczyński, P. Krehlik, A. Czubla, Ł. Buczek, and M. Lipiński," Dissemination of time and RF frequency via a stabilized fibre optic link over a distance of 420 km". Metrologia **50**, 133-145 (2013).
9. M. Lessing, H. S. Margolis, C. T. A. Brown, and G. Marra, " Frequency comb-based time transfer over a 159 km long installed fiber network". Applied Physics Letters **110** (2017)
10. O. Lopez, A. Amy-Klein, C. Daussy, C. Chardonnet, F. Narbonneau, M. Lours, and G. Santarelli," 86-km optical link with a resolution of 2× 10-18 for RF frequency transfer". The European Physical Journal D 48, 35-41 (2008).
11. Whittaker, E. A., Gehrtz, M., & Bjorklund, G. C.," Residual amplitude modulation in laser electro-optic phase modulation". JOSA B, **2**, 1320-1326(1985).
12. F. Frank, F. Stefani, P. Tuckey, and P. E. Pottie," A sub-ps stability time transfer method based on optical modems". IEEE Trans Ultrason Ferroelectr Freq Control **65**, 1001-1006 (2018).
13. Y. Zhi, J. Sun, Y. Zhou, W. Lu, E. Dai, W. Pan and L. Liu," 2.5- and 5-Gbps time-delay selfhomodyne interference differential phase-shift keying optical receiver for space-to-ground communication link". Optical Engineering **58**, 016114 (2019).
14. A. M. J. van Eijk, Y. a. Zhi, J. Sun, E. Dai, Y. Zhou, L. Wang, W. Lu, P. Hou, L. Liu, C. C. Davis, S. M. Hammel, and A. K. Majumdar, "High-data rate differential phase shift keying receiver for satellite-to-ground optical communication link," presented at the Laser Communication and Propagation through the Atmosphere and Oceans2012.
15. L. Yu, R. Wang, L. Lu, Y. Zhu, J. Zheng, C. Wu, B. Zhang, and P. Wang," WDM-based radio frequency dissemination in a tree-topology fiber optic network". Optics Express **23**, 19783-19792 (2015).
16. J. Wang, W. Chen, Q. Liu, N. Cheng, Z. Feng, F. Yang, Y. Gui, and H. Cai," Ultrastable Multiclock Frequency Injection and Dissemination in a Ring Fiber Network". IEEE Photonics Journal **9**, 1-7 (2017).